\documentclass{optica-article}
\journal{oe}
\articletype{Research Article}

\begin{document}
\title{Photon blockade with a trapped $\Lambda$-type three-level atom in asymmetrical cavity}
\author{Xue-Chen Gao,\authormark{1} Xiao-Jie Wu,\authormark{1} Cheng-Hua Bai,\authormark{1} Shao-Xiong Wu,\authormark{1,*} and Chang-shui Yu\authormark{2,\dag}}
\address{\authormark{1} School of Semiconductor and Physics, North University of China, Taiyuan 030051, China\\
\authormark{2} School of Physics, Dalian University of Technology, Dalian 116024, China}
\email{\authormark{*}sxwu@nuc.edu.cn}
\email{\authormark{\dag}ycs@dlut.edu.cn}

\begin{abstract*}
We propose a scheme to manipulate strong and nonreciprocal photon blockades in asymmetrical Fabry-Perot cavity with a $\Lambda$-type three-level atom. Utilizing the mechanisms of both conventional and unconventional blockade, the strong photon blockade is achieved by the anharmonic eigenenergy spectrum brought by $\Lambda$-type atom and the destructive quantum interference effect induced by a microwave field. By optimizing the system parameters, the manipulation of strong photon blockade over a wide range of cavity detuning can be realized. Using spatial symmetry breaking introduced by the asymmetry of cavity, the direction-dependent nonreciprocal photon blockade can be achieved, and the nonreciprocity can reach the maximum at optimal cavity detuning. In particular, manipulating the occurring position of nonreciprocal photon blockade can be implemented by simply adjusting the cavity detuning. Our scheme provides feasible access for generating high-quality nonreciprocal single-photon sources.
\end{abstract*}

\section{Introduction}
The photon blockade (PB) effect \cite{Imamoglu1997} reveals the quantum statistical properties of photons exhibiting antibunching phenomena, where the photons are absorbed or radiated individually one by one, and it can be employed to generate single-photon source. Since generating a single photon is an essential manipulation in quantum computing and quantum simulation, the explorations of the PB effect have been blooming in recent years. In general, there are two different physical mechanisms to explain the PB effect: the conventional photon blockade (CPB) \cite{Imamoglu1997, Hartmann2007} based on the anharmonic eigenenergy spectrum, and the unconventional photon blockade (UPB) \cite{Liew2010, Bamba2011} due to the destructive quantum interference between different transition paths.

For the CPB, it happens in the strong coupling regime between the light and matter, and the harmonicity of the eigenenergy spectrum is broken. Experimentally, starting with the first observation of the CPB effect in the single atom-optical cavity coupled system \cite{Birnbaum2005}, the antibunching phenomenon has been demonstrated in different systems \cite{Faraon2008, Reinhard2012, Hoffman2011, Lang2011, Hamsen2017}. Theoretically, the CPB effect had been investigated widely, such as atom-cavity coupled systems \cite{Dayan2008, Faraon2010}, optomechanical systems \cite{Rabl2011, Liao2013, Lu2013, Lu2015, Xie2016, Wang2015, Zhu2018, Zou2019, Wang2020}, cavity quantum electrodynamic (QED) systems \cite{Ridolfo2012, Bajcsy2013, Wang2017, Radulaski2017, Han2018, Trivedi2019, Hou2019}, waveguide-QED systems \cite{Zheng2011, Zheng2012, Mirza2016}, and circuit-QED systems \cite{Hoffman2011, Liu2014}, and so on.

The strong photon antibunching can also occur in the weak coupling regime; the system must contain multiple degrees of freedom to establish different transition paths, which can be explained utilizing the UPB mechanism. The properties of UPB have also attracted extensive attention. In addition to the experimental demonstration in two coupled superconducting resonators \cite{Vaneph2018}, coupled multimode system \cite{Chakram2022}, and coupled quantum dot-cavity system \cite{Snijders2018}, the UPB effect has also been theoretically studied in coupled two-cavity systems with nonlinear medium \cite{Shen2015, Flayac2016, Zhou2015, Zhou2016, Wang2023}, coupled optomechanical systems \cite{Xu2013, Restrepo2017, Sarma2018, WangDY2020}, coupled quantum dot-cavity systems \cite{Zhang2014, Tang2015, Jabri2022}, quantum-well microcavities system with squeezed light \cite{Jabri2021}, and nonlinear photonic molecule systems \cite{Bamba2011, Xu2014}, and so on.

Combining the PB effect and nonreciprocity, nonreciprocal PB makes the light along one input direction appears bunching phenomenon and shows an antibunching phenomenon from the opposite port. Due to the unidirectional nonclassical property, nonreciprocal PB provides an optional route to implement the nonreciprocal quantum devices, such as single-photon diodes \cite{Xia2014, Scheucher2016, Tang2019}, one-way quantum amplifiers \cite{Abdo2014, Metelmann2015, Malz2018, Shen2018}, and routers of thermal signals \cite{Barzanjeh2018}. Note that nonreciprocal PB has already been theoretically investigated \cite{Huang2018, Li2019, Wang2019, Shen2020, Xu2020, Xia2021, Xia2022, Liu2023, LiuYM2023, Xie2022, Gu2022} and experimentally demonstrated \cite{Yang2023} in recent years. Most of these works focused on utilizing the CPB/UPB mechanism to realize and enhance the PB effect, and a rotating resonator mainly obtains the nonreciprocal PB.

Here, we study such a scenario in which the strong PB and nonreciprocal PB can be generated and controlled with a $\Lambda$-type three-level atom in asymmetrical Fabry-Perot cavity. Firstly, not only  the anharmonicity of the eigenenergy spectrum induced by the $\Lambda$-type atom, but also the additional excitation transition path is introduced through a microwave field, and the strong PB can be achieved by utilizing the mechanism of both CPB and UPB. Secondly, the strong PB can occur in a broad regime of cavity detuning under the selected system parameters and be enhanced under the optimal parameters and the single-excitation resonance condition. Specifically, by taking advantage of the cavity's spatial asymmetrical structure, the strong PB effect can exhibit nonreciprocally with carefully choosing system parameters, i.e., strong antibunching photons can be appeared only in one input port and not vice versa. In particular, the nonreciprocal PB can occur in the very position where we demand it by just changing the cavity detuning. Our scheme has the advantage that it can simultaneously utilize the mechanisms of CPB and UPB to achieve strong PB without additional technologies. In addition, the fragile quantum nonclassical effects are immune to mechanical movement since the system is fixed.

The paper is organized as follows. In Sec. \ref{sec2}, we introduce the model of the system and derive the Hamiltonian. In addition, the optimal parameter condition for strong PB is obtained, and the origin of nonreciprocal PB is discussed. In Sec. \ref{sec3}, we investigate how to optimize the strong PB effect and manipulate the nonreciprocal PB. The conclusion of this work is given in Sec. \ref{sec4}.

\section{Theoretical model}\label{sec2}
\begin{figure}
\centering
\includegraphics[width=0.6\columnwidth]{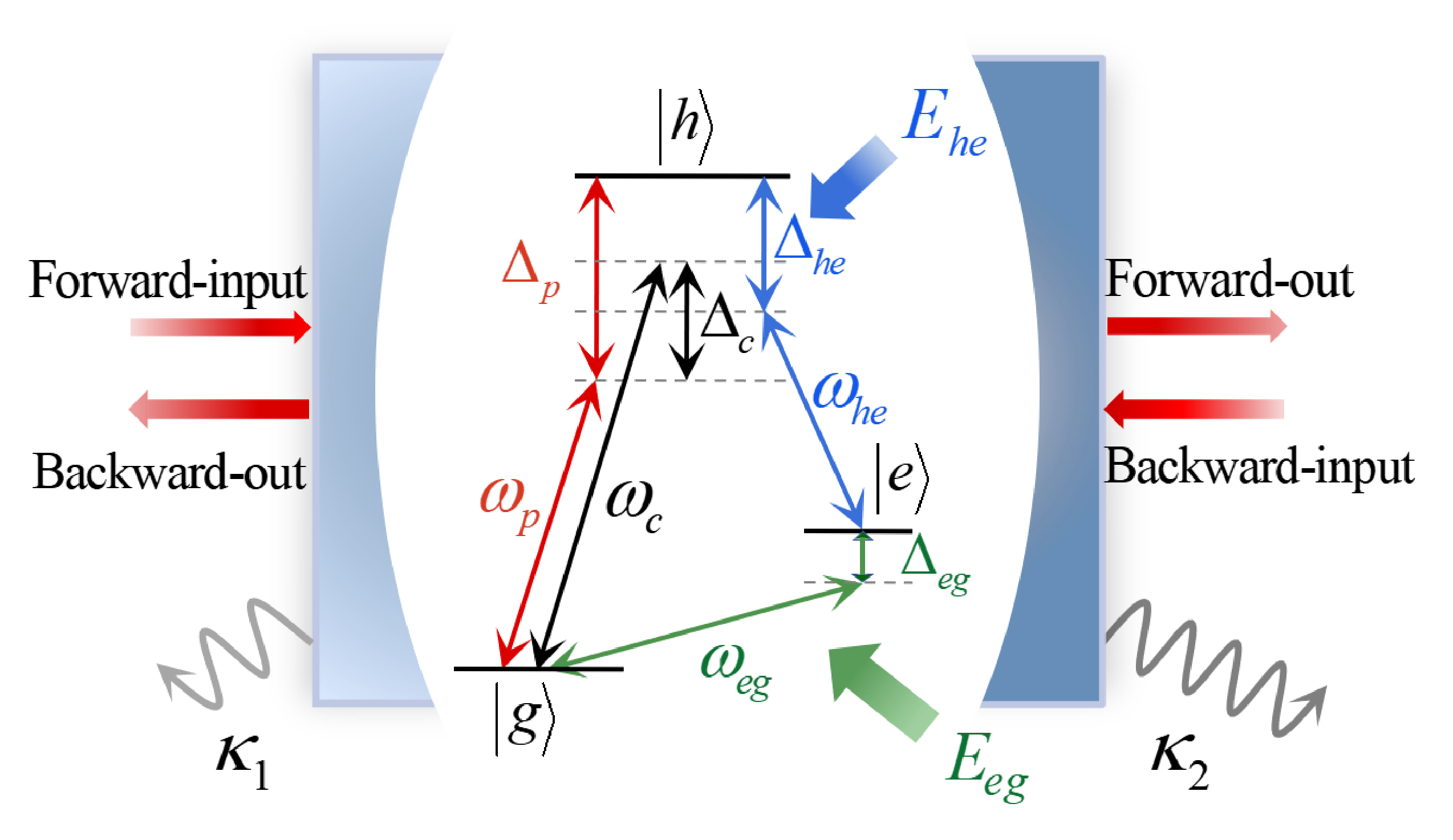}
\caption{Schematic illustration of generation of strong PB and nonreciprocal PB in asymmetrical Fabry-Perot cavity with a $\Lambda$-type three-level atom, which consists of one excited state $|h\rangle$ and two hyperfine ground states $|g\rangle$ and $|e\rangle$. $\kappa_1/\kappa_2$ is the decay rate of the left/right cavity mirror. Three classical coherent fields drive the system; the driving amplitudes, frequencies, and corresponding detunings are given in the text detailed. The case of left input and right output is defined the forward case; otherwise, it is the backward case.}\label{fig:1}
\end{figure}

We consider a $\Lambda$-type three-level atom inside a single-mode asymmetrical Fabry-Perot cavity, as depicted in Fig. \ref{fig:1}. The atom consists of one excited state $|h\rangle$ and two hyperfine ground states $|g\rangle$ and $|e\rangle$ \cite{Mucke2010,Kampschulte2010,You2011,Yang2019}. The asymmetry of the cavity is expressed by the difference of decay rates $\kappa_1$ and $\kappa_2$ for the left and right mirrors, which are determined by the reflectivity of the super-mirrors. The total cavity-decay rate is defined as $\kappa=\kappa_{\text{ave}}+\kappa_{\text{loss}}$, where $\kappa_{\text{ave}}=(\kappa_1+\kappa_2)/2$ is the average cavity-decay rate, and $\kappa_{\text{loss}}$ describes the loss of cavity mode. Since the cavity is high quality ($\kappa_{\text{loss}}\ll\kappa_{\text{ave}}$), the $\kappa_{\text{loss}}$ can be ignored, and $\kappa=\kappa_{\text{ave}}$ can be assumed for simplicity and without loss of generality. The Hamiltonian of the system is written as ($\hbar=1$)
\begin{align}
H_s=&\omega_c{a^\dag}a+\omega_e|e\rangle\langle e|+\omega_h|h\rangle\langle h|\notag\\
&+g({a^\dag}|g\rangle\langle h|+a|h\rangle\langle g|),\label{eq:Hs}
\end{align}
where, the first three terms characterize the free Hamiltonian of cavity field, the energies of atomic states $|e\rangle$ and $|h\rangle$, respectively. Here, $a^\dag(a)$ is the creation (annihilation) operator of the cavity field, and the energy level of ground state $|g\rangle$ is set to be zero and omitted; the last term denotes the interaction between the cavity field and atom with coupling strength $g$.

Three classical coherent fields simultaneously drive the system, and the corresponding driving Hamiltonian is
\begin{align}
H_d=&\Omega e^{i\phi_p}{a^\dag}e^{-i\omega_pt}+E_{he}e^{i\phi_{he}}|h\rangle\langle e|e^{-i\omega_{he}t}\notag\\
&+E_{eg}e^{i\phi_{eg}}|e\rangle\langle g|e^{-i\omega_{eg}t}+\text{H.c.},\label{eq:Hd}
\end{align}
where the first term expresses the cavity mode driven by a weak laser field (Rabi frequency $\Omega$, phase $\phi_p$, frequency $\omega_p$), the second term represents the atomic levels transition $|e\rangle\leftrightarrow|h\rangle$ driven by a $\pi$-polarized classical laser field (amplitude $E_{he}$, phase $\phi_{he}$, frequency $\omega_{he}$), and the third term is a microwave field that drives the atomic transition $|g\rangle\leftrightarrow|e\rangle$ with amplitude $E_{eg}$, phase $\phi_{eg}$, frequency $\omega_{eg}$. The transition between the hyperfine ground states $|g\rangle\leftrightarrow|e\rangle$ is forbidden when the microwave field is absent. H.c. is the abbreviation for ``Hermitian conjugation''.

It is worth noting that due to the inconsistent decay rates of the left and right cavity mirrors, the effective driving strength $\Omega=\sqrt{\kappa_i}b_{\text{in}}(i=1,2)$ of the left and right input port will also be different, where $b_{\text{in}}=\sqrt{P_{\text{in}}/(\hbar\omega_p)}$ represents the driving amplitude under the pump power $P_{\text{in}}$. As shown in Fig. \ref{fig:1}, the laser field entering along the left port of the cavity is defined as the case of forward input; conversely, it is the backward input case when the incident field comes from the right port. We will first discuss the case of forward input, while the case of backward input can be obtained by replacing the decay rate $\kappa_1$ with $\kappa_2$.

The total Hamiltonian of the whole system can be expressed as $H_0=H_s+H_d$. Under the rotating frame with $
U_1=\exp[-i(\omega_p{a^\dag}a+\omega_h|h\rangle\langle h|+\omega_{eg}|e\rangle\langle e|)t]$,
the transformed total Hamiltonian $H_R={U_1^\dag}H_0U_1-i{U_1^\dag}\dot U_1$ can be simplified as
\begin{align}
H_R=&\Delta_c{a^\dag}a+\Delta_{eg}|e\rangle\langle e|+g{a^\dag}|g\rangle\langle h|e^{-i\Delta_pt}\notag\\
&+E_{he}e^{i\phi_{he}}|h\rangle\langle e|e^{i(\Delta_{he}+\Delta_{eg})t}+\Omega e^{i\phi_p}{a^\dag}\notag\\
&+E_{eg}e^{i\phi_{eg}}|e\rangle\langle g|+\text{H.c.},\label{eq:HR}
\end{align}
where $\Delta_c=\omega_c-\omega_p$ is the detuning of the cavity field, $\Delta_{eg}=\omega_e-\omega_{eg}$, $\Delta_p=\omega_h-\omega_p$, $\Delta_{he}=\omega_h-\omega_e-\omega_{he}$ correspond to the detunings of different energy levels of atom, which are illustrated graphically in Fig. \ref{fig:1}.

Under the condition of large detuning, i.e., $|\Delta_p/g|\gg1$ and $|\Delta_{he}/E_{he}|\gg1$, the excited state $|h\rangle$ can be adiabatically eliminated. Therefore, the reduced Hamiltonian with a large detuning approximation can be obtained as
\begin{align}
H_L=&\Delta_c{a^\dag}a+\Delta_e|e\rangle\langle e|-G{a^\dag}a|g\rangle\langle g|-Je^{i\phi_{he}}{a^\dag}|g\rangle\langle e|\notag\\
&+\Omega e^{i\phi_p}{a^\dag}+E_{eg}e^{i\phi_{eg}}|e\rangle\langle g|+\text{H.c.}\label{eq:HL}
\end{align}
For the sake of convenience, the relevant coefficients introduced in $H_L$ (\ref{eq:HL}) are defined as
\begin{align}
\Delta_e=\Delta_{eg}-\frac{E_{he}^2}{\Delta_p},G=\frac{g^2}{\Delta_p},J=\frac{g E_{he}}{\Delta_p},\label{eq:HLcanshu}
\end{align}
where ${\Delta_e}$ denotes the effective detuning of energy level $|e\rangle$, $G$ represents the optical Stark shift, and $J$ is the Raman coupling strength.

To obtain a more simple form of Hamiltonian, a canonical transformation $U_2=\exp(i\phi_p{a^\dag}a-i\phi_{eg}|g\rangle\langle g|)$ on Hamiltonian (\ref{eq:HL}) is performed, and the transformed Hamiltonian $H_T={U_2^\dag}H_LU_2$ can be reorganized as follows
\begin{align}
H_T=&\Delta_c{a^\dag}a+\Delta_e|e\rangle\langle e|-G{a^\dag}a|g\rangle\langle g|-Je^{-i\theta}{a^\dag}|g\rangle\langle e|\notag\\
&+\Omega{a^\dag}+E_{eg}|e\rangle\langle g|+\text{H.c.},\label{eq:HT}
\end{align}
$\theta=\phi_p-\phi_{he}-\phi_{eg}$ is a tunable relative phase among the three classical coherent driving fields.

In the weak-driving limit, i.e., $|\Omega/\kappa|\ll1$ and $|E_{eg}/\kappa|\ll1$, the Hilbert space of a system can be truncated in a low excitation subspace $\{|n,g\rangle,|n-1,e\rangle\}$. Through diagonalizing the transformed Hamiltonian (\ref{eq:HT}), the energy eigenvalues of the system can be obtained
\begin{align}
\varepsilon_{n\pm}=&\frac{(2n-1)\Delta_c+\Delta_e-nG}{2}\notag\\
&\pm\frac{\sqrt{4nJ^2+(nG-\Delta_c+\Delta_e)^2}}{2},\label{eq:varepsilon}
\end{align}
the last term (with $\pm$ symbol) characterizing the upper/lower shifts of energy levels is the origin of the anharmonicity of the eigenenergy spectrum. The Stark shift $G$ not only causes the change of $-nG/2$ on the system energy level but also has an important effect on the degree of energy splitting. The Raman coupling strength $J$ is another parameter that has a significant influence on the energy splitting, and the optimal value of $J$ is related to the decay rate $\kappa_i$ (see the Eq. (\ref{eq:zuiyou})). Therefore, the input direction can also affect the photon's statistical properties.

In the following, we will study the photon blockade effect analytically and numerically calculate the equal-time second-order correlation function \cite{Eleuch2008, Eleuch2009, Jabri2005} $g^{(2)}(0)$ for the asymmetrical cavity mode. To include the influence of system dissipation \cite{Rotter2015, Rotter2009, Eleuch2014}, a phenomenologically imaginary dissipation term is added into the Hamiltonian (\ref{eq:HT}), which can be written as
\begin{align}
H=H_T-i\frac{\kappa}{2}{a^\dag}a,\label{eq:H}
\end{align}
where $\kappa$ is the total decay rate of the cavity. It should be pointed out that the spontaneous emission of the hyperfine ground state $|e\rangle$ has been neglected due to the electric dipole forbidden without the microwave field. In the limit of weak driving, by truncating the Hilbert space of the system up to two-photon excitation subspace $(n\leq2)$, the evolved state can be described as a pure state
\begin{align}
|\psi(t)\rangle=&C_{0g}(t)|0,g\rangle+C_{1g}(t)|1,g\rangle+C_{0e}(t)|0,e\rangle\notag\\
&+C_{2g}(t)|2,g\rangle+C_{1e}(t)|1,e\rangle,\label{eq:psi}
\end{align}
where the coefficients $C_{n\alpha}(t)(\alpha=g,e)$ are the probability amplitudes of states $|n,\alpha\rangle$. Based on the Schr\"{o}dinger equation $i|\dot{\psi}(t)\rangle=H|\psi(t)\rangle$, the dynamical equations for the coefficients $C_{n\alpha}(t)$ are governed by the differential equations
\begin{align}
i{\dot C}_{1g}=&\sqrt{\kappa_1}b_{\text{in}}C_{0g}+MC_{1g}+\sqrt{2\kappa_1}b_{\text{in}}C_{2g}\notag\\
&-Je^{-i\theta}C_{0e}+E_{eg}C_{1e},\notag\\
i{\dot C}_{2g}=&\sqrt{2\kappa_1}b_{\text{in}}C_{1g}+2MC_{2g}-\sqrt{2}Je^{-i\theta}C_{1e},\notag\\
i{\dot C}_{0e}=&E_{eg}C_{0g}-Je^{i\theta}C_{1g}+\Delta_e C_{0e}+\sqrt{\kappa_1}b_{\text{in}}C_{1e},\notag\\
i{\dot C}_{1e}=&E_{eg}C_{1g}-\sqrt{2}Je^{i\theta}C_{2g}+\sqrt{\kappa_1}b_{\text{in}}C_{0e}+NC_{1e},\notag\\
i{\dot C}_{0g}=&\sqrt{\kappa_1}b_{\text{in}}C_{1g}+E_{eg}C_{0e},\label{eq:gailv}
\end{align}
where, the variables $M=\Delta_c-i\frac{\kappa}{2}-G$ and $N=\Delta_c-i\frac{\kappa}{2}+\Delta_e$ are introduced for simplicity.

Under the weak-driving limit, it is reasonable to assume that the state $|0,g\rangle$ does not evolve and $C_{0g}\simeq1$. On the one hand, the evolved state $|\psi(t)\rangle$ (\ref{eq:psi}) can be obtained by solving the differential equations of Eq. (\ref{eq:gailv}) numerically through the Runge-Kutta method \cite{Xia2021, Xia2022}, and will become steady-state when the evolution time is long enough. On the other hand, the steady-state solution of probability amplitudes $C_{n\alpha}$ can be obtained analytically by the perturbation method, i.e., discarding the higher-order terms for the lower-order variables in Eq. (\ref{eq:gailv}), and the solutions are
\begin{align}
C_{1g}\simeq&\frac{E_{eg}Je^{-i\theta}+\sqrt{\kappa_1}b_{\text{in}}\Delta_e}{J^2-M\Delta_e},\notag\\
C_{0e}\simeq&\frac{E_{eg}M+\sqrt{\kappa_1}b_{\text{in}}Je^{i\theta}}{J^2-M\Delta_e},\notag\\
C_{2g}\simeq&\frac{(E_{eg}Je^{-i\theta}+\sqrt{\kappa_1}b_{\text{in}}N)C_{1g}+\sqrt{\kappa_1}b_{\text{in}}Je^{-i\theta}C_{0e}}{\sqrt2(J^2-MN)},\notag\\
C_{1e}\simeq&\frac{(E_{eg}M+\sqrt{\kappa_1}b_{\text{in}}Je^{i\theta})C_{1g}+\sqrt{\kappa_1}b_{\text{in}}MC_{0e}}{J^2-MN}.\label{eq:gailvjie}
\end{align}

According to the quantum input-output theory \cite{Gardiner1985} and introducing the output field operator $b_{\text{out}}=-i\sqrt{\kappa_2}a$, the equal-time second-order correlation function of the output photon can be expressed as
\begin{align}
g^{(2)}(0)=\frac{\langle b_{\text{out}}^\dag b_{\text{out}}^\dag b_{\text{out}} b_{\text{out}}\rangle}{\langle b_{\text{out}}^\dag b_{\text{out}}\rangle^2}=\frac{2P_2}{(P_1+2P_2)^2},\label{eq:g20}
\end{align}
where the probability distributions of photon number are
\begin{align}
P_1=&(|C_{1g}|^2+|C_{1e}|^2)/\mathcal{N},\notag\\
P_2=&|C_{2g}|^2/\mathcal{N} \label{eq:P}
\end{align}
with the omitted normalization constant $\mathcal{N}=|C_{0g}|^2+|C_{1g}|^2+|C_{0e}|^2+|C_{2g}|^2+|C_{1e}|^2\approx 1$.

To achieve a perfect single-photon blockade, i.e., $g^{(2)}(0)\sim0$, the excitation probability $P_2$ of two-photon state $|2,g\rangle$ is required to be 0. Through straightforward derivation, the explicit expression is tedious and not shown here; however, the system parameters satisfy the following optimal relation
\begin{align}
Je^{-i\theta}=-\frac{b_{\text{in}}\sqrt{\kappa_1}N(E_{eg}Je^{-i\theta}+b_{\text{in}}\sqrt{\kappa_1}\Delta_e)}
{J(E_{eg}^2e^{-i\theta}+b_{\text{in}}^2{\kappa_1}e^{i\theta})+b_{\text{in}}\sqrt{\kappa_1}E_{eg}(M+\Delta_e)}.
\label{eq:zuiyou}
\end{align}
The inconsistency between the decay rate $\kappa_1$ and $\kappa_2$ will lead to the discrepancy of optimal parameters $J$ and $\theta$ for the forward and backward input case, which means that the position of perfect photon blockade occurred for the forward and backward input is different in the $J$-$\theta$ parameter phase space when other system parameters are constant. Therefore, one can realize a unidirectional antibunching by tuning the Raman coupling strength $J$ and relative phase $\theta$, which indicates that nonreciprocal photon blockade can be manipulated on demand. A detailed analysis is given in the next section.

\section{Strong PB and Nonreciprocal PB}\label{sec3}
\subsection{Strong PB}
\begin{figure}[h]
\centering
\includegraphics[width=0.4\columnwidth]{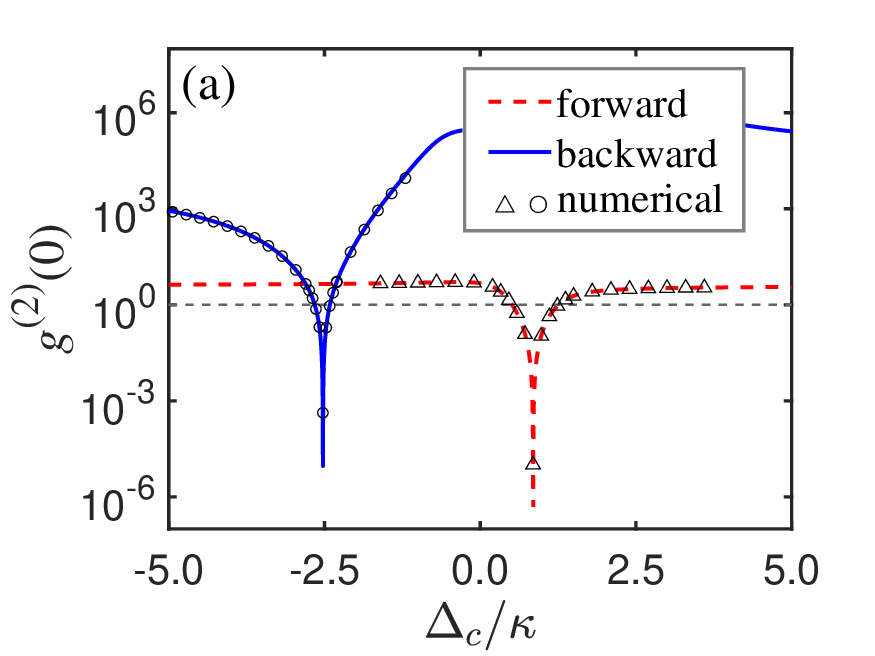}
\includegraphics[width=0.4\columnwidth]{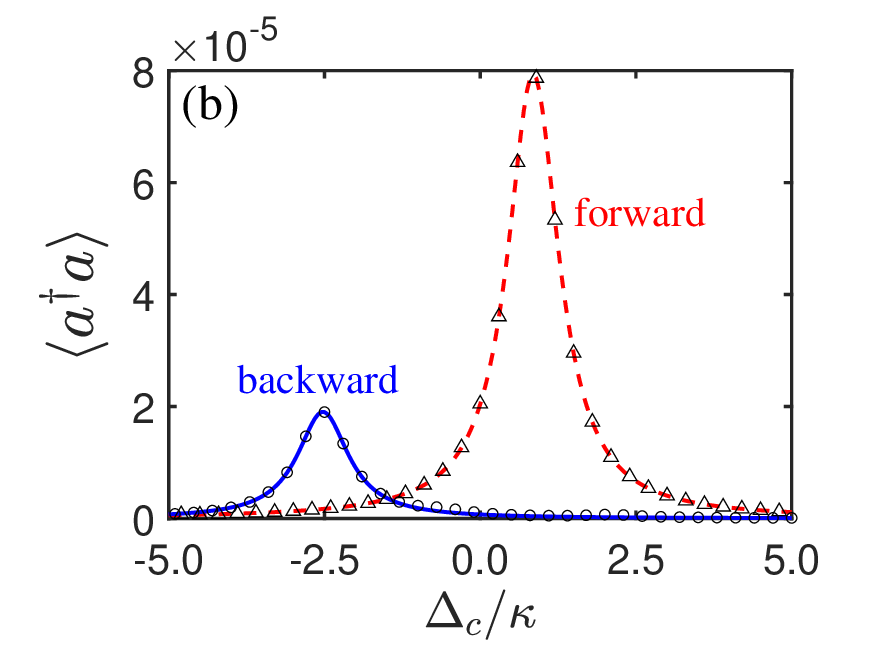}
\caption{(a) The second-order correlation function $g^{(2)}(0)$ and (b) intracavity photon number $\langle a^{\dag}a \rangle$ as a function of cavity-pumping detuning $\Delta_c$ for optimal relation $J$ and $\theta$ under the strength of driving light $b_{\text{in}}/\sqrt{\kappa}=0.02$. The dashed red and solid blue lines represent the analytical solutions of forward ($\kappa_1/\kappa=0.2$) and backward ($\kappa_2/\kappa=1.8$) incident light, respectively. Meanwhile, the black triangular and circular points are numerical results based on the Runge-Kutta method.}\label{fig:2}
\end{figure}

According to the current experimental techniques \cite{Volz2011,Li2017,Sato2012,Spillane2005,Yang2019}, the system parameters are set as: $\kappa=2\pi\text{MHz}$, $\kappa_1/\kappa=0.2$, $\kappa_2/\kappa=1.8$, $g/\kappa=10$, which accesses to the strong coupling regime. The external driving fields are chosen appropriately: $\Delta_p/\kappa=100$, $\Delta_e/\kappa=-0.5$, $b_{\text{in}}/\sqrt{\kappa}=0.02$, and $E_{eg}/\kappa=0.01$, which satisfy the conditions of large detuning and weak driving. Comparing with the  Cesium D2 transition between $6~^2\text{S}_{1/2}\leftrightarrow6~^2\text{P}_{3/2}$, the pumping of the cavity can be chosen as $\lambda=852\text{nm}$ continuous wave tunable laser uner the power $1.16\text{fW}$, and the hyperfine ground state $\text{F=3}$ and $\text{F}=4$ can be split about $50\text{KHz}$ by a $9.2\text{GHz}$ microwave filed under the power about $10\text{W}$, the power of atomic driving laser is tuned dependent on the system parameters \cite{Yang2019, Yang2023}. Combining the optimal parameter relation in Eq. (\ref{eq:zuiyou}) with the optimal cavity-pumping detuning $\Delta_c^{\text{opt}}=G+J^{2}/\Delta_e$ (notice that $\Delta_e$ cannot be zero) under the single-excitation resonance condition, we will analyze the influence of system parameters on the relationship between the PB effect and maximum single-photon occupancy probability, which are displayed in Fig. \ref{fig:2}(a) and (b), respectively. The dashed red line (forward case) and the solid blue line (backward case) are obtained based on the analytical steady-state result; meanwhile, the black triangular and circular points are obtained through the numerical simulation by the Runge-Kutta method, and there is an excellent agreement between the analytical result and the numerical method. Due to the reliability of the analytical steady-state solution under the weak driving condition, we will investigate the PB effects based on the analytical result in the following work.

In Fig. \ref{fig:2}(a), we plot the equal-time second-order correlation function $g^{(2)}(0)$ as a function of cavity-pumping detuning $\Delta_c$ under the optimal parameter condition (\ref{eq:zuiyou}). When the cavity driving laser with Rabi frequency $\Omega$ is input from the left or right port, the curves of $g^{(2)}(0)$ for both the forward and backward cases show a sharp decline, and the sliding position is precisely located at the optimal cavity-pumping detuning, which indicates strong photon blockade occurring. In addition, owing to the dip of $g^{(2)}(0)$ for the forward and the backward cases are split distinctly by the different cavity detuning $\Delta_c$, and the $g^{(2)}(0)$ for the opposite case is greater than 1; this feature provides an platform to investigate the nonreciprocal photon blockade effect, which will be discussed in subsection 3.2.

To confirm that the maximum single-photon occupation probability happens in the optimal cavity-pumping detuning condition, the variation of intracavity photon number $\langle a^{\dag}a \rangle\simeq|C_{1g}|^2$ as a function of cavity-pumping detuning $\Delta_c$ for the forward and backward cases are displayed in Fig. \ref{fig:2}(b). Both the curves of $\langle a^{\dag}a\rangle$ for the forward (dashed red line) and backward (solid blue line) reach the maximum at the corresponding optimal cavity-pumping detuning $\Delta_c^{\text{opt}}$, which indicates that the probability of single photon occupancy reaches maximum and is correspondents to the strong photon antibunching occurring. Obviously, we can obtain the strong photon blockade and high single-photon occupation probability at the same time.

\begin{figure}[b]
\centering
\includegraphics[width=0.4\columnwidth]{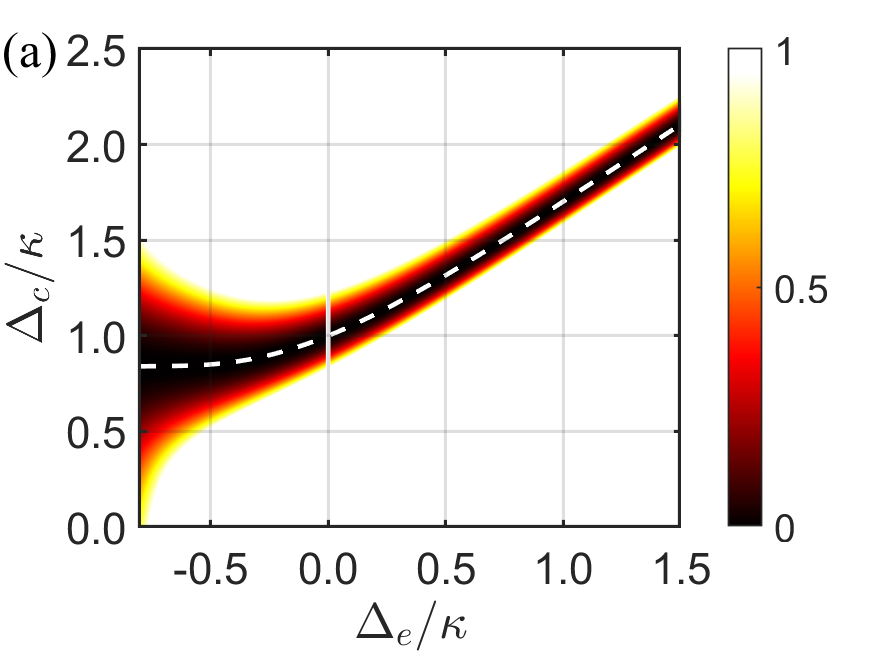}
\includegraphics[width=0.4\columnwidth]{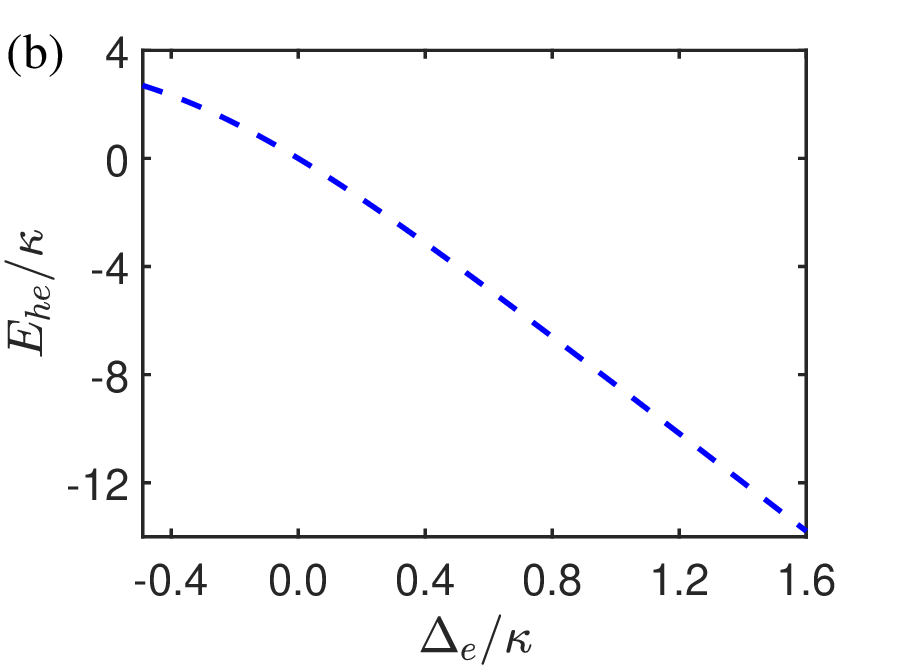}
\includegraphics[width=0.4\columnwidth]{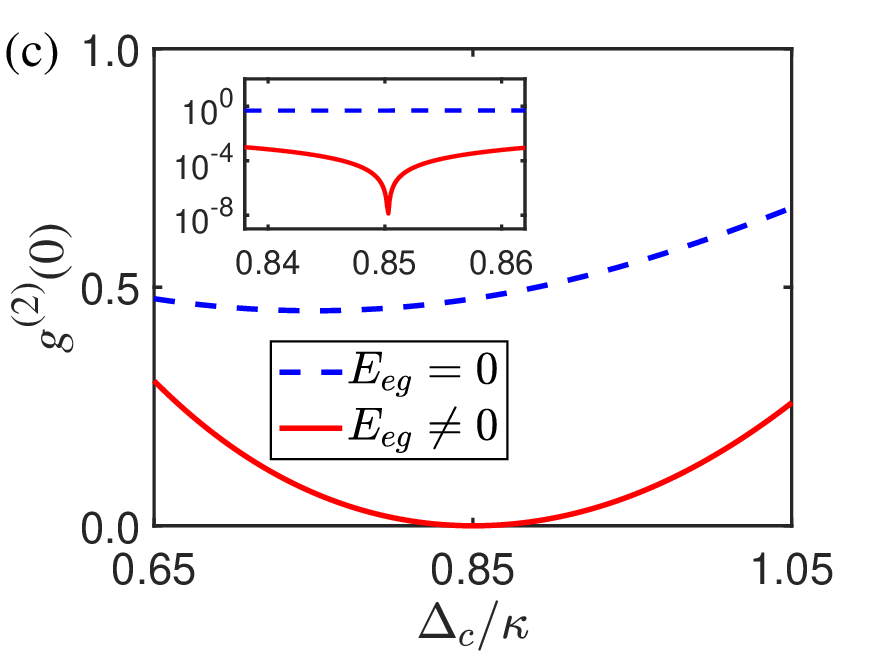}
\includegraphics[width=0.4\columnwidth]{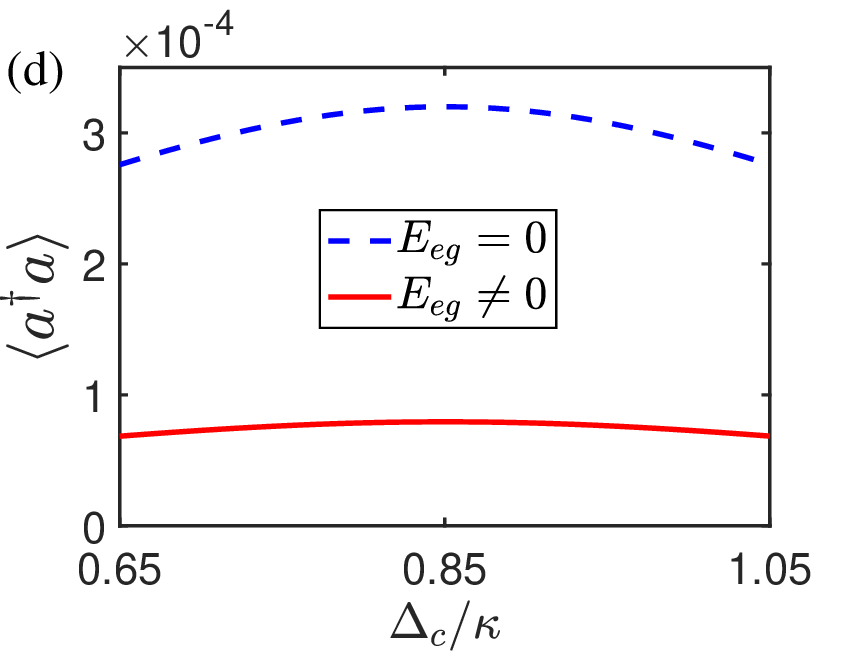}
\caption{(a) The second-order correlation function $g^{(2)}(0)$ versus the detunings $\Delta_e$ and $\Delta_c$. The dashed white line represents the strong photon blockade under the optimal relation. In panel (b), the optimal driving amplitude $E_{he}$ versus the detuning $\Delta_e$ is shown. Panels (c) $g^{(2)}(0)$ and (d) intracavity photon number $\langle a^{\dag}a \rangle$ as a function of cavity detuning $\Delta_c$ with the microwave field present (solid red line) or absent (dashed blue line). In all panels, $\kappa_1/\kappa=0.2$, $J$ and $\theta$ are set the optimal relation.}\label{fig:3}
\end{figure}

To gain more insight into photon blockade, we will take the total cavity-decay rate $\kappa$ as the energy unit, fix the large detuning $\Delta_p/\kappa=100$, the driving strength of input field $b_{\text{in}}/\sqrt{\kappa}=0.02$, and discuss the system parameters on the PB effects. In Fig. \ref{fig:3}(a), we draw the variation of $g^{(2)}(0)$ in the detuning parameter $\Delta_e$-$\Delta_c$ phase space, where the dashed white line indicates the occurrence of strong photon blockade (the minimum of PB) under optimal parameter condition (\ref{eq:zuiyou}). We can find that the position of the strong photon blockade that appears in the cavity detuning $\Delta_c$ is changed along with the effective detuning $\Delta_e$, and this trend is well consistent with the dashed white line. Noticing that the strong photon blockade cannot be occurred in $\Delta_e/\kappa=0$ according to the expression of optimal cavity-pumping detuning $\Delta_c^{\text{opt}}=G+J^2/\Delta_e$. We can appropriately adjust the effective detuning $\Delta_e$ to ensure that the optimal photon blockade appears within a wide range of cavity detuning $\Delta_c$. Combing with the optimal parameter relation in Eq. (\ref{eq:zuiyou}) and the Raman coupling strength $J$ in Eq. (\ref{eq:HLcanshu}), we can realize the adjustment of effective detuning $\Delta_e$ by only tuning the driving amplitude $E_{he}$, for which we plot the relation between the driving amplitude $E_{he}$ and the effective detuning $\Delta_e$ in Fig. \ref{fig:3}(b). It is worth mentioned that since the transition $|h\rangle\leftrightarrow|e\rangle$ is suppressed under the condition of large detuning $\Delta_{he}\gg E_{he}$, the driving amplitude $E_{he}$ does not need to meet the weak-driving condition.

The influence of microwave field on the generation of strong photon blockade is studied, and the results are shown in Fig. \ref{fig:3}(c). One can find that when the microwave field is absent (dashed blue line), although the equal-time second-order correlation function $g^{(2)}(0)$ can still be less than 1, the photon blockade effect is not ideal. This is because the optimal parameter relation for $|C_{2g}|=0$ can be simplified as $J^2+(\Delta_c-i\kappa/2+\Delta_e)=0$ when $E_{eg}=0$, which means that we cannot get a solution of this equation in field of real number. Accordingly, we cannot obtain a satisfactory photon blockade effect. Crucially, when the microwave driving field is present (solid red line), one can find that $g^{(2)}(0)\simeq0$ appears near the detuning $\Delta_c/\kappa=0.85$, in which case a quasi-perfect photon blockade effect can be achieved without resorting to the other external auxiliary. At the same time, we can also observe that the occupancy probability of single photon in the cavity is inevitably suppressed when the microwave field exists, which is shown in Fig. \ref{fig:3}(d). Based on the above analysis, we can conclude that the presence of microwave field can indeed improve the photon blockade effect in the cost of sacrificing the intracavity photon number.

\begin{figure}[h]
\centering
\includegraphics[width=0.6\columnwidth]{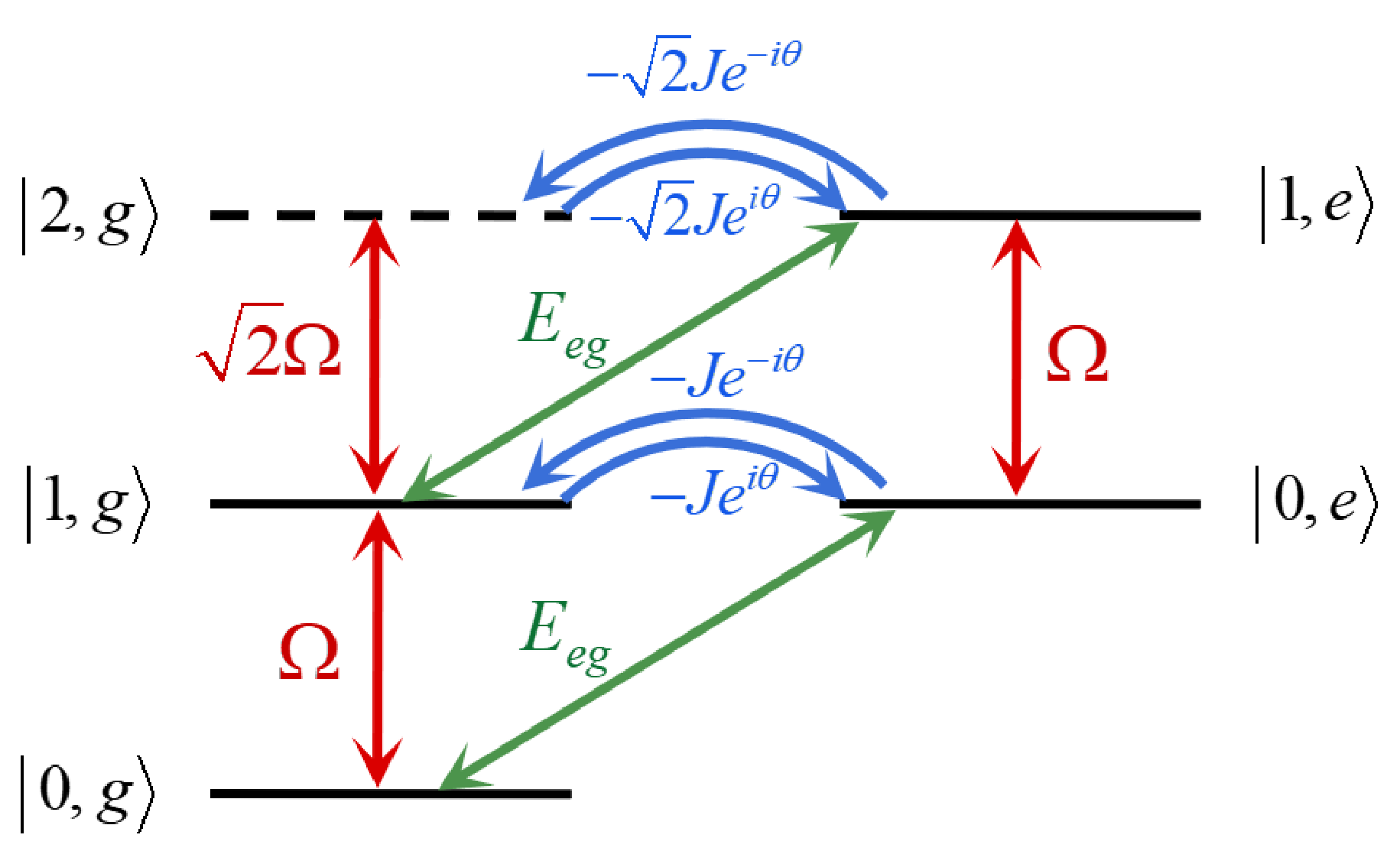}
\caption{The quantum transition paths of different energy levels. The destructive quantum interference effect can eliminate the two-photon excited state $|2,g\rangle$ (dashed black line).}\label{fig:4}
\end{figure}

In order to reveal the physical mechanism behind the phenomena, the quantum transition paths between different energy levels are plotted in Fig. \ref{fig:4}. It should be emphasized that the microwave field can enhance the photon blockade because of the destructive quantum interference effect between different quantum transition paths. Specifically, the interference paths occur between the transitions $|1,g\rangle \xrightarrow{\sqrt2\Omega} |2,g\rangle$ induced by the cavity driving field and $|1,g\rangle \xrightarrow {E_{eg}} |1,e\rangle \xrightarrow{-\sqrt2Je^{-i\theta}} |2,g\rangle$ induced by the combination of both the microwave field and atomic driving field. On the other hand, it is worth mentioning that due to the destructive quantum interference effect between the transition paths $|0,g\rangle \xrightarrow {\Omega} |1,g\rangle$ and $|0,g\rangle \xrightarrow {E_{eg}} |0,e\rangle \xrightarrow {-Je^{-i\theta}} |1,g\rangle$, the occupancy probability of single photon in the cavity will also be inevitably suppressed. Even the number of single photons in the cavity can be suppressed, the microwave field is indispensable to  enhance the PB effect through quantum interference processes.

\begin{figure}
\centering
\includegraphics[width=0.4\columnwidth]{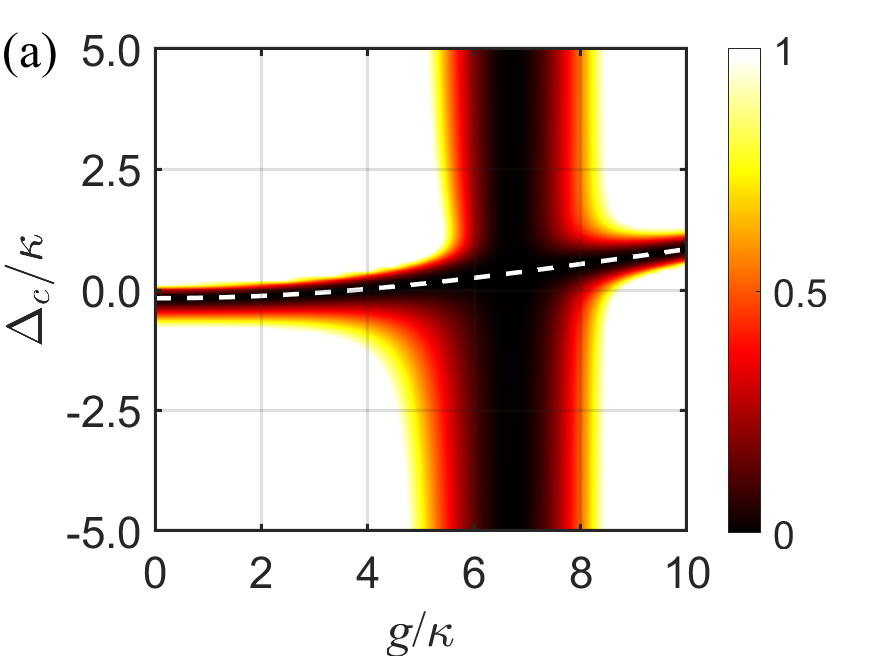}
\includegraphics[width=0.4\columnwidth]{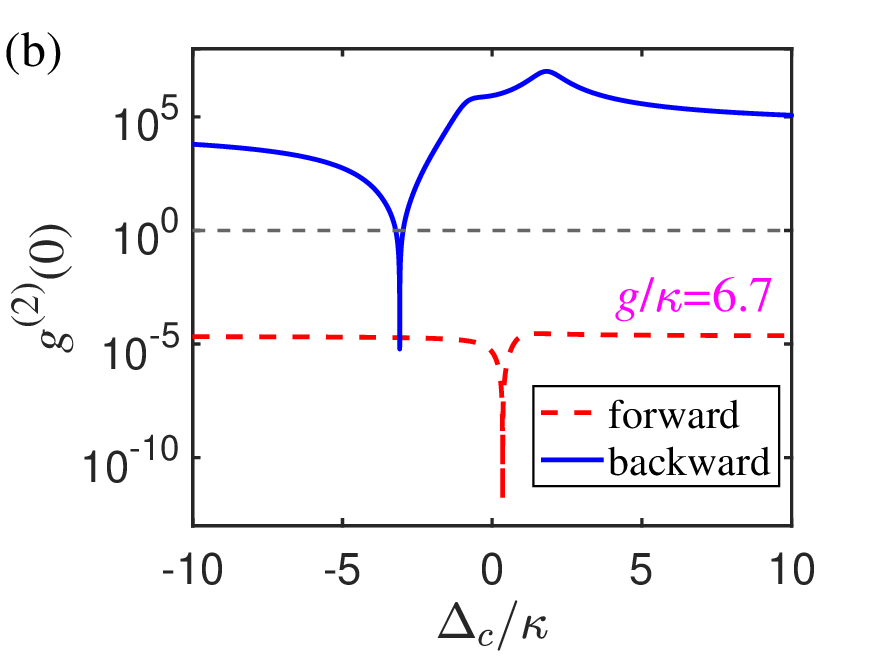}
\includegraphics[width=0.4\columnwidth]{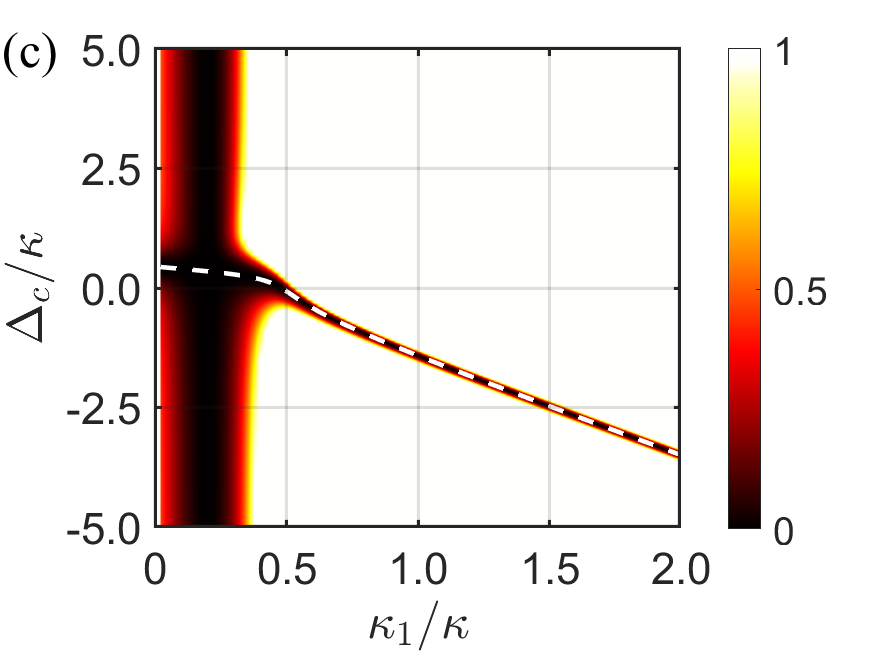}
\caption{(a) The second-order correlation function $g^{(2)}(0)$ versus $g$ and $\Delta_c$. The dashed white line corresponds to the relation of the optimal blockade position. Panel (b) shows $g^{(2)}(0)$ as a function of cavity detuning $\Delta_c$ when the cavity-atom coupling strength is $g/\kappa=6.7$. Panel (c) is $g^{(2)}(0)$ versus the cavity-decay rate $\kappa_1$ and cavity detuning $\Delta_c$ for $g/\kappa=6.7$. In all panels, $J$ and $\theta$ are given by the optimal relation in Eq. (\ref{eq:zuiyou}), the other parameters are the same as those used in Fig. \ref{fig:2}.}\label{fig:5}
\end{figure}

More importantly, in order to investigate the effect of cavity-atom coupling strength $g$ on the photon blockade, the changes of $g^{(2)}(0)$ in the $g$-$\Delta_c$ parameter phase space are shown in Fig. \ref{fig:5}(a). The dashed white line denotes the relation of optimal blockade position versus the cavity-atom coupling strength $g$. As the cavity-atom coupling strength $g$ increasing, the position of the strong photon blockade on the cavity detuning $\Delta_c$ will be changed accordingly, and the variation trend is in good agreement with the white dashed line. Meanwhile, we also observed a sudden change in the vicinity of $g/\kappa=6.7$. To gain more insight into PB about this sudden change point, we display $g^{(2)}(0)$ as a function of cavity detuning $\Delta_c$ in the condition of cavity-atom coupling strength $g/\kappa=6.7$ in Fig. \ref{fig:5}(b). Remarkably, one can observe that the equal-time second-order correlation function $g^{(2)}(0)$ for the forward case (dashed red line) is much smaller than 1 within the entire range of cavity detuning $\Delta_c$, and there is a perfect photon blockade occurring in the position of optimal detuning $\Delta_c^{\text{opt}}$. This means we can achieve a satisfactory photon blockade effect in a vast cavity detuning $\Delta_c$ with a specific cavity-atom coupling parameter $g$. While, for the backward case (solid blue line), the photon blockade only occurs in a reasonably narrow regime.  To further explain this phenomenon, we plot the variation of $g^{(2)}(0)$ for the forward case in the $\kappa_1$-$\Delta_c$ parameter phase space with $g/\kappa=6.7$ in Fig. \ref{fig:5}(c). One can find that the regime of strong photon blockade for $\kappa_1/\kappa>0.4$ has a similar trend as the dashed white line. However, in the range of $\kappa_1/\kappa<0.4$, the equal-time second-order correlation function $g^{(2)}(0)$ is less than 1 in all chosen cavity detuning ranges. One reasonable explanation of this phenomenon may be that the Rabi frequency $\Omega=\sqrt{\kappa_i}b_{\text{in}}$ is very small when the cavity-decay rate $\kappa_1$ is small enough, and the intracavity photons cannot be excited to higher energy levels. The blockade effect can occur throughout the detuning range.

\subsection{Nonreciprocal PB}
\begin{figure}[b]
\centering
\includegraphics[width=0.4\columnwidth]{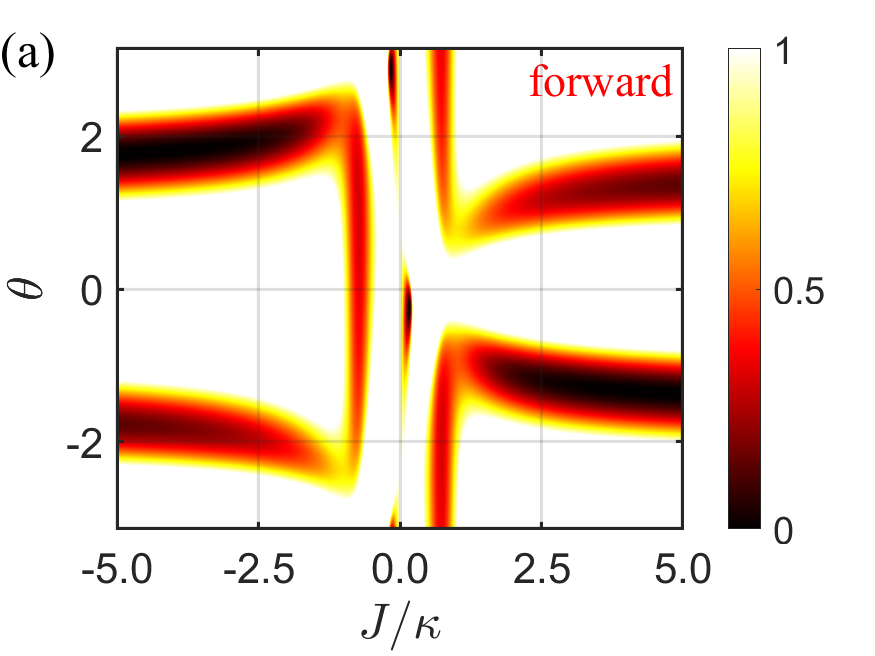}
\includegraphics[width=0.4\columnwidth]{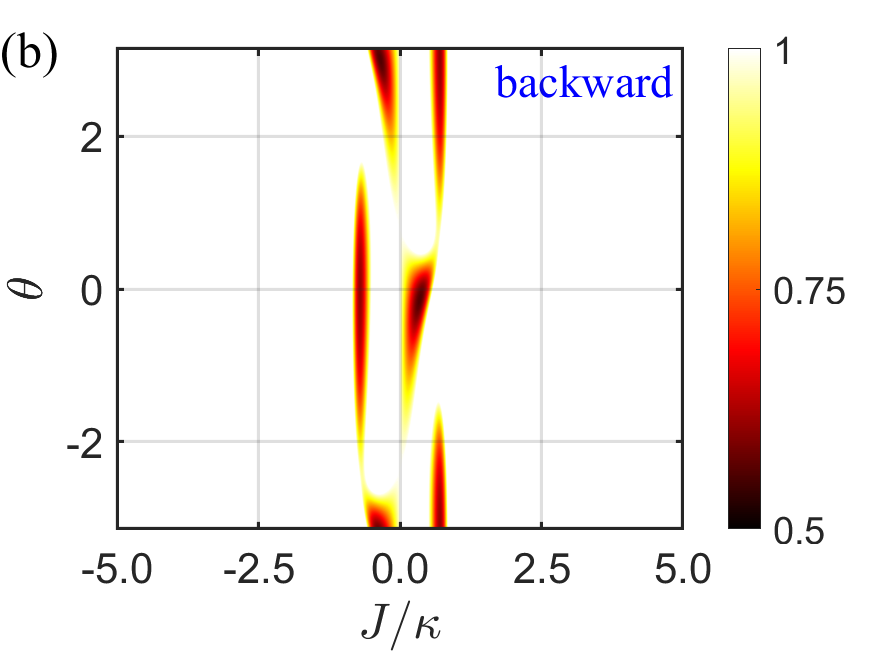}
\includegraphics[width=0.4\columnwidth]{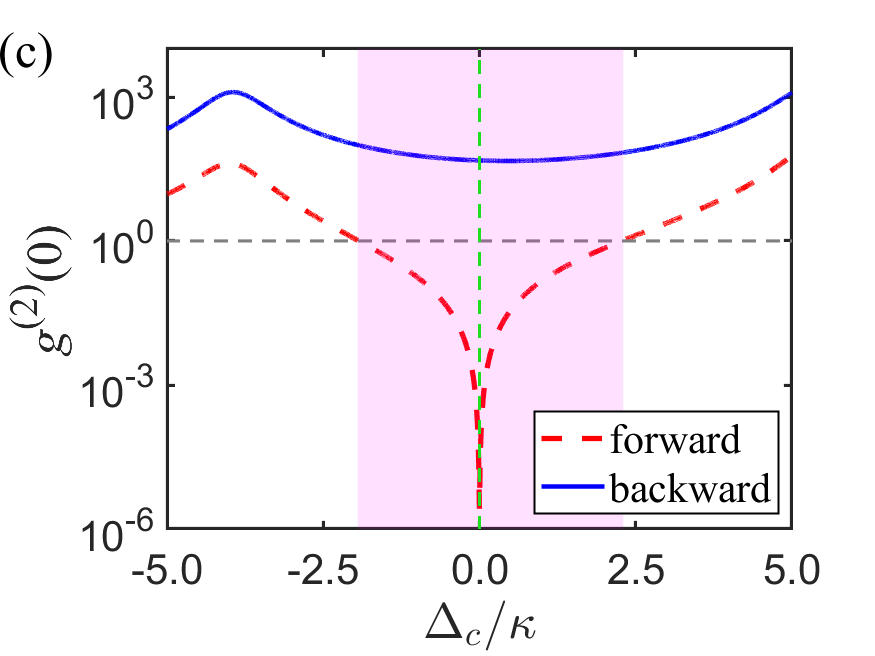}
\includegraphics[width=0.4\columnwidth]{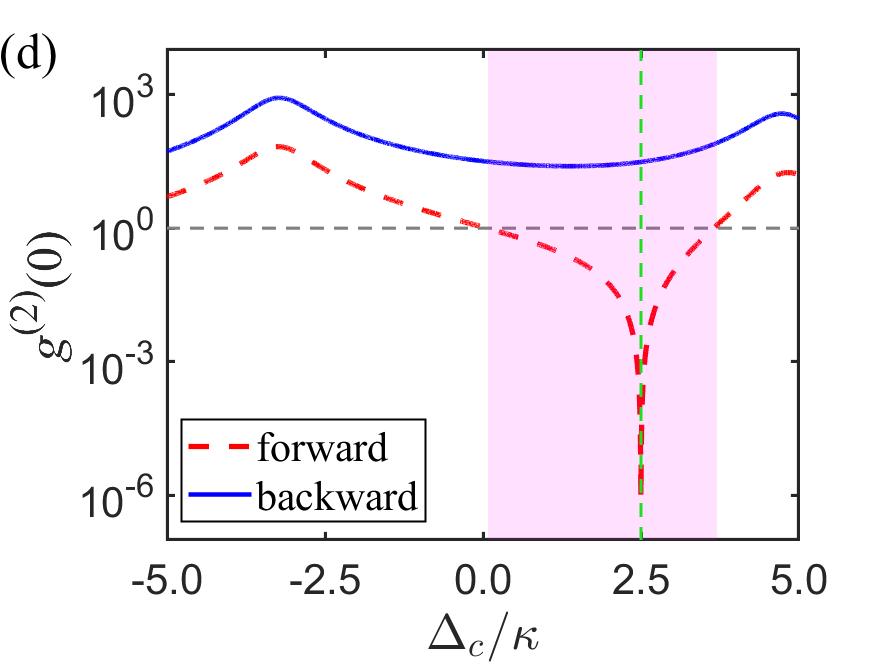}
\caption{Panels (a) and (b) are the second-order correlation function $g^{(2)}(0)$ versus the Raman coupling strength $J$ and relative phase $\theta$ for the forward case and backward case with $\Delta_c/\kappa=0$, respectively. The nonreciprocal photon blockade can be realized with suitable $\theta$ in the ranges $J/\kappa<-1.5$ or $J/\kappa>1.5$. (c) $g^{(2)}(0)$ versus the cavity detuning $\Delta_c$ for the forward and backward cases. The pink region denotes that the nonreciprocal photon blockade can be occurred, and the dashed green line represents the location of the maximum nonreciprocal photon blockade. Panel (d) is an example of manipulating the maximum nonreciprocal photon blockade at an on-demand position, such as the $\Delta_c/\kappa=2.5$. The system parameters are the same as those used in Fig. \ref{fig:2}.}\label{fig:6}
\end{figure}

Inspired by the above analysis, we can employ the spatial symmetry breaking induced by asymmetrical cavities to generate and manipulate the nonreciprocal photon blockades. Specifically, due to the asymmetrical decay rate $\kappa_i$ of the cavity's left and right mirror, the Rabi frequency ($\Omega=\sqrt{\kappa_i}b_{\text{in}}$) will be changed with the input port, which leads to distinct photons statistical properties. On contrary to Fig. \ref{fig:2}(a), where the optimal parameters $J$ and $\theta$ is adopted to calculate the equal-time second-order correlation function $g^{(2)}(0)$ for the forward and backward cases, the system parameters (including $J$ and $\theta$) remain unchanged to catch the prominent nonreciprocity feature of photon blockade (for example $g^{(2)}(0)\ll1$ for the forward case and $g^{(2)}(0)>1$ for the backward case). To achieve this purpose, we plot the variation of the equal-time second-order correlation function $g^{(2)}(0)$ for the forward and backward cases in the $J$-$\theta$ parameter phase space in Fig. \ref{fig:6}(a) and (b), respectively. It is convinced that $g^{(2)}(0)$ is much less than 1 for the forward case under suitable relative phase $\theta$ when the Raman coupling strength $J$ is in the ranges of $J/\kappa<-1.5$ or $J/\kappa>1.5$, while $g^{(2)}(0)$ for the backward case is greater than 1 in the all ranges of relative phases. Therefore, the nonreciprocal photon blockade can be realized by jointly adjusting the Raman coupling strength and relative phase, which can be carried out by changing the amplitude of atomic driving $E_{he}$, the atomic detuning $\Delta_p$, and the relative phases of the three classical coherent fields.

Next, we will optimize the nonreciprocal photon blockade, which are shown by the pink region in Fig. \ref{fig:6} (c) and (d). Though selecting the minimum value of equal-time second-order correlation function $g^{(2)}(0)$ within the range of $J/\kappa<-1.5$ or $J/\kappa>1.5$, and taking the Raman coupling strength $J$ and relative phase $\theta$ as the corresponding optimal relation, the photon blockade can display the property of significant nonreciprocity. To prove the validity of our analysis, we plot $g^{(2)}(0)$ as a function of cavity detuning $\Delta_c$ in Fig. \ref{fig:6}(c), where the values of $J$ and $\theta$ are selected based on the above analysis. It is vividly shown that $g^{(2)}(0)\ll1$ (dashed red line) for the forward case and $g^{(2)}(0)>1$ (solid blue line) for the backward case in the pink-covered region, which confirms the nonreciprocity of photon blockade. For the forward case, one can notice that the value of $g^{(2)}(0)$ reaches the minimum at the detuning $\Delta_c/\kappa=0$ (the dashed green line), which indicates the location of maximum nonreciprocal photon blockade occurring.

In addition, we also give an example to manipulate the maximum nonreciprocal photon blockade, which can be realized by adjusting the cavity detuning $\Delta_c$ and shown in Fig. \ref{fig:6}(d). By jointly adjusting the system parameters, we can make the maximum nonreciprocal photon blockade occurs at a specific position, such as $\Delta_c/\kappa=2.5$. The results accurately validate the feasibility of our analysis. Thus, we can carry out not only the manipulation of strong photon blockade but also nonreciprocal blockades.

\section{Conclusion}\label{sec4}
In conclusion, we have proposed a simple scheme to implement and manipulate the strong and nonreciprocal photon blockade with a trapped $\Lambda$-type three-level atom in asymmetrical Fabry-Perot cavity. The anharmonic eigenenergy spectrum and the destructive quantum interference effect between different transition paths mainly induce the PB effect. For strong PB, the occurring position of PB can be manipulated by changing the driving amplitude $E_{he}$, and the microwave field can introduce an additional destructive quantum interference path, which can enhance the blockade effect greatly. Moreover, the PB can occur in a broad detuning range under suitable cavity-atom coupling strength. For the nonreciprocal PB, it can be generated by utilizing the intrinsic spatial symmetry breaking of the asymmetrical cavity, maximizing and manipulating the occurring position by jointly adjusting the system parameters. Our work provides a novel, experimentally feasible plan to achieve both strong and nonreciprocal photon blockades, and generate high-quality nonreciprocal single-photon sources.

\begin{backmatter}
\bmsection{Funding}
National Natural Science Foundation of China (12175029, 12204440), Fundamental Research Program of Shanxi Province (20210302123063, 202103021223184).

\bmsection{Disclosures}
The authors declare no conflicts of interest.

\bmsection{Data Availability}
Data underlying the results presented in this paper are not publicly available at this time but may be obtained from the authors upon reasonable request.
\end{backmatter}

\newcommand{\doi}[2]{\href{https://doi.org/#1}{\color{blue}#2}}

\end{document}